# Monitoring the Digital Divide

E. Canessa, H. A. Cerdeira
*The Abdus Salam ICTP, Trieste 34100, Italy*
ejds@ictp.trieste.it

W. Matthews, R. L. Cottrell
*SLAC, Stanford, CA 94025, USA*
{warrenm,cottrell}@slac.stanford.edu

It is increasingly important to support the large numbers of scientists working in remote areas and having low-bandwidth access to the Internet. This will continue to be the case for years to come since there is evidence from PingER performance measurements that the, so-called, digital divide is not decreasing. In this work, we review the collaborative work of The Abdus Salam International Center for Theoretical Physics (ICTP) in Trieste -a leading organization promoting science dissemination in the developing world- and SLAC in Stanford, to monitor by PingER, Universities and Research Institutions all over the developing world following the recent *"Recommendations of Trieste"* to help bridge the digital divide. As a result, PingER's deployment now covers the real-time monitoring of worldwide Internet performance and, in particular, West and Central Africa for the first time. We report on the results from the ICTP sites and quantitatively identify regions with poor performance, identify trends, discuss experiences and future work.

## 1. OVERVIEW

A large community of scientists from developing countries cannot or can only partially participate or benefit from electronic science due to the lack of adequate network capacity or performance and awareness about alternatives. This can adversely affect both individual scientists and large international collaborations such as those in High Energy and Nuclear Physics (HENP), where typically about 10% of the collaborating sites are in developing countries.

To assist in making information available to scientists worldwide a multidisciplinary group of international experts gathered for an open round table on *"Developing Country Access to On-line Scientific Publishing: Sustainable Alternatives"* in October 2002 at the Abdus Salam International Centre for Theoretical Physics (ICTP) in Trieste, Italy [1]. The meeting was sponsored by ICSU, IUPAP, UNESCO, TWAS and WIF. Among the 10 recommendations made [2], one was specific to monitoring: *"To devote resources to monitor in real time the connectivity of research and educational institutions in developing countries and to encourage (and devote resources to) the development of the connectivity"*.

In December 2002 a letter was sent to the ICTP electronic Journals Distribution Service (eJDS) collaborators [3] with the statement *"To improve the effectiveness of the eJDS and make a survey of those places around the World, which have the need of Internet infrastructure; we plan to monitor by PingER, Universities and Research Institutions all over the developing world"*. As a result, the PingER/eJDS project is now monitoring hosts at sites in over 40 countries subscribing to the eJDS service.

In this work we report on how well PingER/eJDS's deployment now covers the monitoring of worldwide Internet performance, report on the results and quantitatively identify regions with poor performance, including real-time monitoring of some African sites for the first time.

## 2. PingER/eJDS MONITORING

Researchers in the world's poorest nations, where Internet connections can be slow or prohibitively expensive, can now receive some scientific papers free of charge via the e-mail based eJDS system [3]. Via the eJDS, developing world scientists can now have access to a much wider range of current scientific information and findings than ever before. The eJDS procedure to follow is similar to that used when connected to any Web server by selecting (hyper)links [4, 5].

Publishers are now able to reach scientists who would otherwise not have either the technical or financial means to read articles in their eJournals in a timely fashion. The Abdus Salam ICTP can broaden its vital role to meet its mandate and transfer knowledge to scientists in the developing countries.

To improve the effectiveness of the eJDS, the plans are to extend its reach, by providing support for setting up the main access to the Internet for remote campuses. To decide where and how best to provide support to campuses of remote Universities and Research Institutions in the developing world, it is necessary to quantify first the current performance and we plan to do this by PingER monitoring [6].

PingER (*Ping End-to-end Reporting*) is the name given to the Internet End-to-end Performance Measurement (IEPM) project to monitor end-to-end performance of Internet links (see also [7]). This project, initially set up to monitor connectivity to high energy and nuclear physics institutions in many countries, has grown to monitor many other scientific collaborations and now monitors sites in over 75 countries that between them have over 99% of the worldwide users of the Internet.

To measure network performances, the standard ICMP echo [8] based Internet ping facility is used. PingER packet loss rate has been found to be a good measure of the quality of links with loss rates over 0.1%, thus the reports in this paper are for packet loss rates.

At losses of 4-6% or more video-conferencing becomes irritating and non-native language speakers become unable to communicate. The occurrence of long delays of 4 seconds (such as may be caused by timeouts in recovering from packet loss) or more at a frequency of 4-5% or more is also irritating for interactive activities such as telnet and X windows. Conventional wisdom among TCP researchers holds that a loss rate of 5% has a significant adverse effect on TCP performance, because it will greatly limit the size of the congestion window and hence the transfer rate, while 3% is often substantially less serious. A random loss of 2.5% will result in Voice Over Internet Protocols (VOIP) becoming slightly annoying every 30 seconds or so. A more realistic burst loss pattern will result in VOIP distortion going from not annoying to slightly annoying when the loss goes from 0 to 1%. Since TCP throughput goes as $1/\sqrt{(loss)}$ [9], it is important to keep losses low for achieving high throughput. To assist in categorizing the losses PingER defines the following loss quality categories: $< 0.1\%$ = excellent; $>= 0.1\%$ and $< 1\%$ = good; $>= 1\%$ and $< 2.5\%$ = acceptable; $>= 2.5\%$ and $< 5\%$ = poor; $>= 5\%$ and $< 12\%$ = very poor; $> 12\%$ bad.

From January 2003, the PingER project has been extended to African countries and other developing countries in collaboration with the eJDS project. This has successfully provided outreach beyond high energy nuclear and particle physics. Special attention is drawn to academic sites with low-bandwidth connections to the outside world.

This active PingER/eJDS monitoring is essential to catalogue critical needs for networking infrastructure, to understand real performance, set expectations, identify problem areas, provide information for troubleshooting, and to rationalize the allocation of (financial, hardware, human, *etc*) resources to improve the Quality of Service and performances.

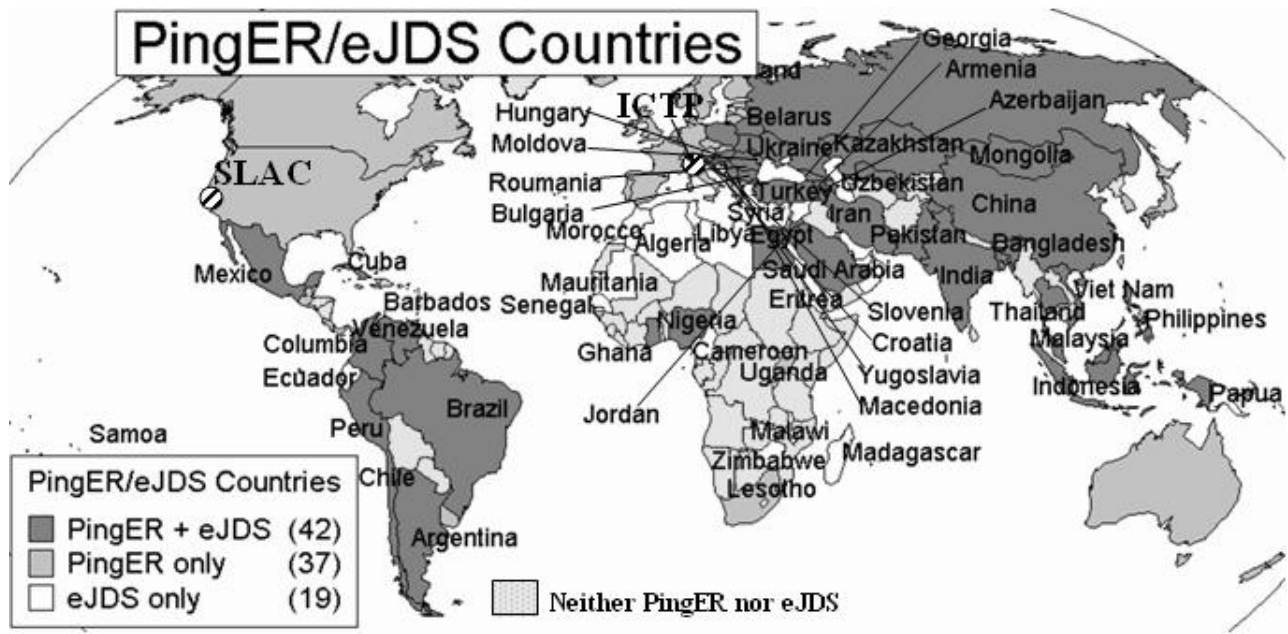

Figure 1: Countries with sites being monitored by PingER and subscribing to eJDS (dark grey), being measured by PingER but not subscribing to eJDS (light grey), subscribing to eJDS only (white), and other countries stippled.

## 3. PRELIMINARY PERFORMANCE RESULTS

The impact on the remotely monitored host is minimal, no software needs to be installed or maintained, no special account is required, and the extra traffic is limited to accommodating 10 pings each with 100Bytes every 30 minutes, *i.e.*, per direction, 8 kbits/s for 10 seconds per half hour or about 5bits/s per monitor host – remote host pair on average. The designated host needs to be available 24 hours/day, 365 days/year, apart from occasional outages. It must also be able to respond to pings from the Internet (*e.g.,* ping must not be blocked or rate limited).

In Fig.1 we show a world map identifying countries with hosts monitored by PingER and with institutes subscribing to the eJDS service. Countries with a dark grey background have institutes participating in the PingER/eJDS monitoring. Countries with a light grey background have hosts monitored by PingER but the sites are not subscribing to eJDS. Countries with a white background have institutes subscribing to eJDS but are not monitored yet. The numbers in parentheses are the number of countries in the relevant category. Currently PingER measurements are made from SLAC to over 40 countries with sites subscribing to the eJDS service.

PingER/eJDS packet loss measurements reported here mainly cover from October 2002 until the end of February 2003.

### 3.1. Middle East

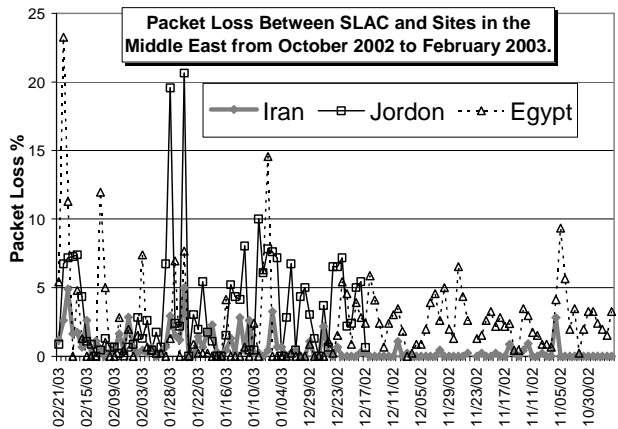

**Figure 2: Packet Loss in percent between SLAC and sites in the Middle East.**

Performance to sites in the Middle East shown in Fig.2 is mostly variable and indicative of congested links. Jordan (average loss ~ 3.6%) and Egypt (average loss ~ 2.5%) have poor performance. Some sites perform well. In particular, some sites in Iran (average loss ~ 0.6%) exhibit low losses consistent with adequately provisioned links. Israel in general seems to be well connected since we observe very low packet loss to sites in Israel (average loss ~ 0.06%) compared to other countries in the region. Saudi Arabia (the site is an oil platform connected via a satellite) exhibits the worst losses (average loss ~ 5.6%). The Round Trip Times (RTT) to Saudi Arabia (not shown here) are also characteristic (*i.e.,* RTTs of one to two seconds) of an overloaded satellite link.

### 3.2. Caucasus & Central Asia

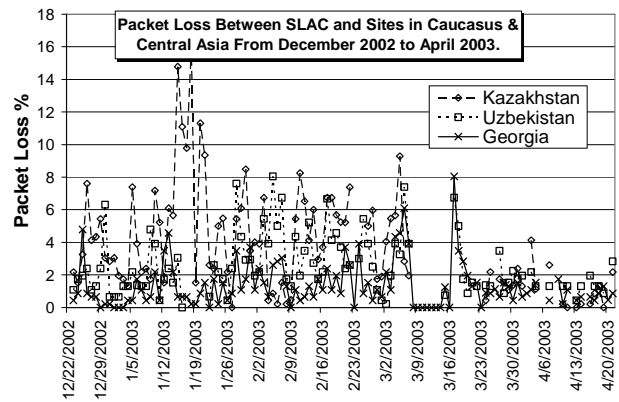

**Figure 3: Packet Loss in percent between SLAC and sites in the Caucasus and Central Asia.**

Performance to sites in the Caucasus and Central Asia is shown in Fig. 3. The link to Kazakhstan goes via Stockholm and Moscow utilizing land lines with an RTT of about 300ms. Losses to Kazakhstan are poor to bad sometimes exceeding over 12% and a median for this period of over 3%, and with much variability and big differences (factor of 3.5) between weekday and weekend losses indicating congestive losses. Both Georgia and Uzbekistan are routed via an earth station located at DESY in Hamburg, Germany and then via a satellite to the relevant country. Both Georgia and Uzbekistan have RTTs of about 700ms. Median losses to Uzbekistan are about 2% and to Georgia about 1%, less than those to Kazakhstan even though the RTT to Kazakhstan is less. The variability of the losses to Georgia (standard deviation of ~ 1.4 %) and Uzbekistan (standard deviation ~ 1.8 %) are also less than those to Kazakhstan (standard deviation ~3.2 %)

Initiatives such as the silk-road project are expected to improve the performance to this region [10].
.

### 3.3. Latin America

Performance to sites in Latin America is variable as can be seen in Fig. 4. Sites in countries with well organized research networks and a connection to the AMPATH backbone [11] perform well (average packet loss ~ 1%) for Argentina, Brazil and Chile. Countries with well organized research networks but not yet connected to or using AMPATH (*e.g.,* Uruguay and Venezuela) have similar loss rates but much more variability (the standard deviation is about three times as large). Sites like those in Guatemala and Peru (average loss ~ 4.3%), that route to SLAC across commercial networks perform less well with large (average loss ~ 4.3%), variable packet loss. The graph shows packet loss between SLAC and eJDS subscribers grouped by country. These measurements indicate some eJDS subscribers (*e.g.,* Uruguay and Venezuela) could achieve improved performance by connecting to AMPATH.

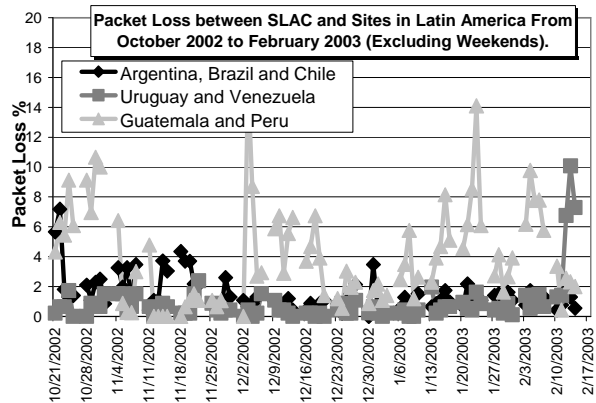

**Figure 4:** Packet Loss in percent between SLAC and sites in Latin America (excluding weekends).

### 3.4. Asia

Examples of packet loss between SLAC and sites in South and East Asia shown in Fig. 5 and 6, respectively, follow a similar pattern to Latin America. Certain better connected countries, (*e.g.,* Japan average loss for this period ~ 0.06% and Taiwan average loss ~ 0.1%) perform very well, but other locations, such as Bangladesh, Indonesia, Malaysia and Pakistan (all over ~ 3% average loss) and India and Thailand (around ~ 2% average loss) perform less well. Fig. 6 shows low packet loss between SLAC and eJDS subscribers in Japan but higher and more variable loss to sites in China (average loss ~ 1.2%) and Korea (average loss ~ 0.7%). The Asia-Pacific advanced network (APAN) provides excellent connectivity to this region [12]. Studies continue in understanding routing and bottlenecks.

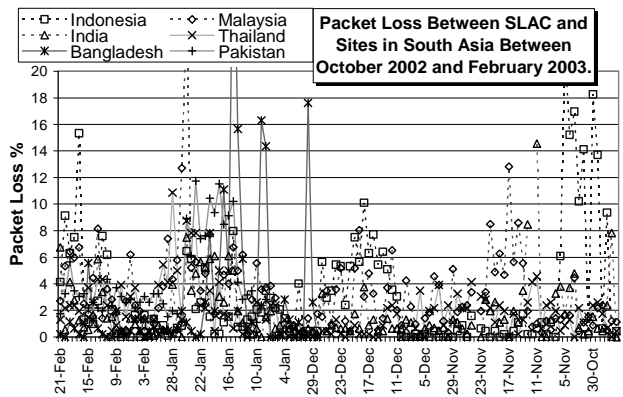

**Figure 5:** Packet Loss in percent between SLAC and sites in South Asia.

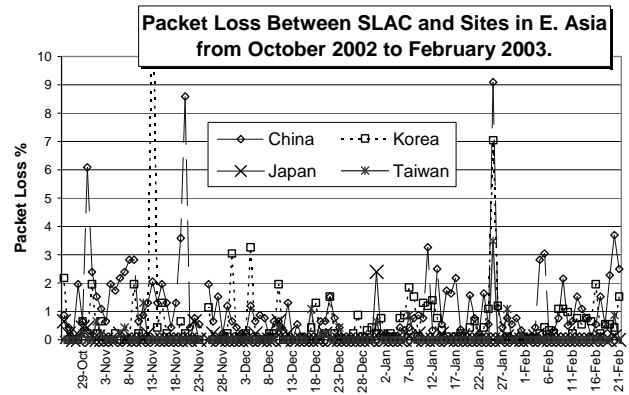

**Figure 6:** Packet Loss in percent between SLAC and sites in East Asia.

### 3.5. West and Central Africa

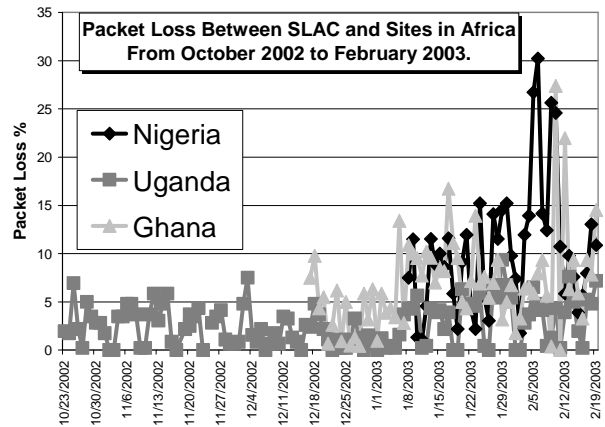

**Figure 7:** Packet Loss in percent between SLAC and sites in West Africa.

PingER/eJDS sites in West and Central Africa are routed to SLAC by satellite connections. Such connectivity results in very long RTTs (*e.g.,* 1500ms for Nigeria), and impacts interactive applications. Fig. 7 shows packet loss between SLAC and eJDS sites in each of three countries being monitored: Nigeria (average loss ~ 10%), Uganda (average loss 3%) and Ghana (average loss ~ 7%). In all cases the high packet loss during the working week and the lower packet loss at weekends is highly indicative of congested links, although the RTT for Uganda is found to be quite stable. Perhaps this is due to rate limiting or the presence of only a few users. There are a number of projects that try to assist in improving performance to sites in Africa [13].

### 3.6. Trends

Fig. 8 shows exponential fits to the monthly median TCP throughputs estimated using the Mathis formula [9] and the PingER RTTs and losses measured from hosts in Energy Sciences Network (ESnet) Laboratories in the U.S. to hosts in various regions of the world. The measurements go back several years, in some cases as far as January 1995. The numbers in parentheses are the number of monitoring host - remote host pairs included in the fitted data. The line labeled Edu refers to .edu sites which are generally associated with U.S. educational institutions. The dots are for a performance increasing by 80%/year or a factor of 10 in 4 years.

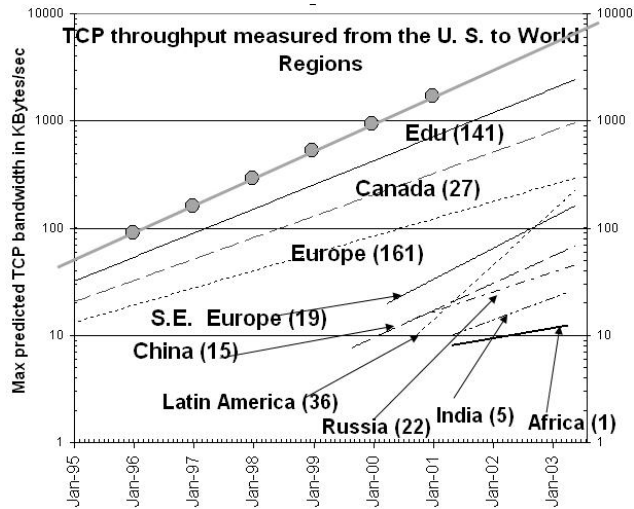

**Figure 8: TCP throughput measured from the U.S. to various regions of the world.**

It can be seen that though regions such as Latin America and S. E. Europe were several years behind Europe, Canada and Edu sites, they are catching up. Other regions such as China, Russia and India are also many years behind and are not catching up. Africa (note the report only shows one country, Uganda, which is the only country for which we currently we have long term measurements) appears to be falling even further behind.

## 4. CONCLUDING REMARKS

PingER/eJDS deployment now covers the real-time monitoring of worldwide Internet performance and, in particular, West and Central Africa for the first time. Table 1 shows a summary of the losses and RTTs to developing countries in various regions of the world. It can be seen that the regions with the worst performance (measured by packet loss) are Central Asia, Africa, South Asia and the Middle East (excepting Israel), all of which have losses that are poor to bad.

Table 1: Packet losses and RTTs for representative developing countries, as seen from SLAC, February 2003

| Region | Countries | Median packet loss | Median RTT (ms) |
|---|---|---|---|
| South Asia | Bangladesh, India, Indonesia, Malaysia, Pakistan, Thailand, Vietnam | 4.5% | 674 |
| E. Asia | China, Korea, Mongolia, Singapore | 1.1% | 263 |
| Central Asia | Kazakhstan, Uzbekistan | 12% | 542 |
| Caucasus | Georgia | 1.6% | 720 |
| Middle East | Egypt, Iran, Jordan, Saudi Arabia, Turkey | 3.4% | 566 |
| Africa | Ghana, Nigeria, & Uganda | 6% | 930 |
| Latin America | Argentina & Brazil | 1.9% | 263 |
| Latin America | Uruguay & Venezuela | 0.3% | 277 |
| Latin America | Guatemala & Peru | 1.5% | 407 |

From these preliminary measurements, there is evidence that some developing nations are many years behind and are not catching up the exponential growth of Internet performance in industrialized nations. This discrepancy in performance is exemplified by the recent end-to-end network throughput record set between California and Switzerland by a team of researchers from Caltech, CERN, SLAC and LANL [14]. This achieved a sustained rate of 2.36 Gbits/s (or ~ 300 MBytes/s) or over a TByte/hour. This is over 3000 times the performance between SLAC in California and sites in developing countries. Even if one takes production paths available today between academic and research sites in the U.S., Europe, and Japan one can regularly achieve over 200-300 Mbits/s [15] or over 1000 times that available to digital divide countries.

It is necessary to advertise both the eJDS and the PingER/eJDS Monitoring Project among Scientists in remote areas. Furnished with the PingER measurements it is possible to better understand network performance, set expectations for the performance of interactive applications, and decide how to allocate resources.

These monitoring efforts, using open source technologies, are a good example to help quantify the digital divide. They can provide valuable information to compare the current performance to industrialized countries with those to less developed countries and also to look at the relative rates of improvement.

They also allow us to identify regions that have poor access and need improvement to bring them up to an acceptable level, and in some cases even suggest ways in which this may be facilitated.

The present measurements and reports provide information and a challenge to all NGO/Agencies working on, or financing, projects to bridge the digital divide.

## Acknowledgments

The authors wish to thank all volunteer scientists at remote hosts participating in the eJDS and PingER/eJDS monitoring projects.

This work was supported in part by the Director, Office of Science, Office of Advanced Scientific Computing Research, Mathematical, Information, and Computational Sciences Division under the U.S. Department of Energy. The SLAC work is under Contract No. DE-AC03-76SF00515.